\documentclass[conference]{IEEEtran}
\IEEEoverridecommandlockouts
\usepackage{cite}
\usepackage{float}
\usepackage{amsmath,amssymb,amsfonts}
\usepackage{graphicx}
\usepackage{textcomp}
\usepackage{xcolor}
\usepackage{acronym}
\usepackage{algorithm}
\usepackage{algorithmic}
\usepackage{siunitx}  
\usepackage{multirow}
\usepackage{bm} 
\usepackage{tikz}
\usepackage{pgfplots}
\usetikzlibrary{matrix}
\usetikzlibrary{positioning}
\usepackage{makecell}
\usepackage{subcaption}
\pgfplotsset{compat=1.18}
\usepackage{subcaption} % better than subfigure
\usetikzlibrary{patterns}
\usepackage[inline]{enumitem}
\usetikzlibrary{patterns.meta}
\usepackage[colorlinks=true,linkcolor=black,anchorcolor=black,citecolor=black,filecolor=black,menucolor=black,runcolor=black,urlcolor=black]{hyperref}
\usepackage{cleveref}

\def\BibTeX{{\rm B\kern-.05em{\sc i\kern-.025em b}\kern-.08em
    T\kern-.1667em\lower.7ex\hbox{E}\kern-.125emX}}

\begin{document}

%\title{Joint Scheduling of a Transparent 5G Bridge in a Time-Aware TSN System}
\title{Joint Resource Allocation to Transparently Integrate 5G TDD Uplink with Time-Aware TSN}
\author{\IEEEauthorblockN{Laura Becker, Yash Deshpande, Wolfgang Kellerer\\}
\IEEEauthorblockA{Chair of Communication Networks, Technical University of Munich, Germany \\
Email: \{laura.alexandra.becker, yash.deshpande,  wolfgang.kellerer\}@tum.de}}

\maketitle

\acrodef{AF}{Application Function}
\acrodef{BAT}{Burst Arrival Time}
\acrodef{BE}{Best Effort}
\acrodef{BD}{Bridge Delay}
\acrodef{BS}{Burst Size}
\acrodef{BSR}{Buffer Status Report}
\acrodef{CBS}{Committed Burst Size}
\acrodef{CIR}{Committed Information Rate}
\acrodef{CG}{Configured Grant}
\acrodef{CQF}{Cyclic Queuing and Forwarding}
\acrodef{CQI}{Channel Quality Indicator}
\acrodef{CNC}{Centralized Network Configuration}
\acrodef{COT}{Commercial Off-The-Shelf}
\acrodef{CUC}{Centralized User Configuration}
\acrodef{DC-GBR}{Delay Critical GBR}
\acrodef{DL}{Downlink}
\acrodef{DS-TT}{Device Side TSN Translator}
\acrodef{DRR}{Deficit Round Robin}
\acrodef{EBS}{Excess Burst Rate}
\acrodef{EDF}{Earliest Deadline First}
\acrodef{EIR}{Excess Information Rate}
\acrodef{GCE}{Gate Control Entry}
\acrodef{GCL}{Gate Control List}
\acrodef{IPV}{Internal Priority Value}
\acrodef{MAC}{Media Access Control}
\acrodef{MCS}{Modulation and Coding Scheme}
\acrodef{MDBV}{Maximum Data Burst Volume}
\acrodef{NR}{New Radio}
\acrodef{NW-TT}{Network Side TSN Translator}
\acrodef{OFDM}{Orthogonal Frequency Division Multiplexing}
\acrodef{PDB}{Packet Delay Budget}
\acrodef{PCP}{Priority Code Point}
\acrodef{PER}{Packet Error Rate}
\acrodef{PF}{Proportional Fair}
\acrodef{PDU}{Protocol Data Unit}
\acrodef{PSFP}{Per-Stream Filtering and Policing}
\acrodef{PUCCH}{Physcial Uplink Control Channel}
\acrodef{QoS}{Quality of Service}
\acrodef{RAN}{Radio Access Network}
\acrodef{RE}{Resource Element}
\acrodef{RB}{Resource Block}
\acrodef{RLC}{Radio Link Control}
\acrodef{RRC}{Radio Resource Control}
\acrodef{SDAP}{Service Data Adaptation Protocol}
\acrodef{SDU}{Service Data Unit}
\acrodef{SPS}{Semi-Persistent Scheduling}
\acrodef{SR}{Scheduling Request}
\acrodef{TAS}{Time-Aware Shaper}
\acrodef{TBS}{Transport Block Size}
\acrodef{TC}{Traffic Class}
\acrodefplural{TC}[TCs]{Traffic Classes}
\acrodef{TDD}{Time Division Duplex}
\acrodef{TSCAI}{Time Sensitive Communication Assistance Information}
\acrodef{TSCAC}{Time Sensitive Communication Assistance Container}
\acrodef{TSN}{Time-Sensitive Networking}
\acrodef{TSpec}{Traffic Specification}
\acrodef{TT}{TSN Translator}
\acrodef{UE}{User Equipment}
\acrodef{UL}{Uplink}
\acrodef{UPF}{User Plane Function}
\acrodef{URLLC}{Ultra Reliable Low Latency Communication}
\acrodef{UTC}{Coordinated Universal Time}
\acrodef{gNB}{Next Generation Node B}

\begin{abstract}
To enable mobility in industrial communication systems, the seamless integration of 5G with \ac{TSN} is a promising approach. Deterministic communication across heterogeneous 5G-TSN systems requires joint scheduling between both domains. A key prerequisite for time-aware end-to-end scheduling is determining the forwarding delay for each \ac{TSN} \acl{TC} at every bridge, referred to as \ac{BD}. Hence, to integrate 5G as a transparent \ac{TSN} bridge, the 5G \ac{BD} must be determined and guaranteed.
Unlike wired bridges, the 5G \ac{BD} relies on wireless resource management characteristics, such as the \acl{TDD} pattern and radio resource allocation procedure. In particular, traditional \ac{UL} schedulers are optimized for throughput but often fail to meet the deadline requirements.

To address this challenge, we propose a heterogeneous radio resource scheduler that integrates static and dynamic scheduling. 
The algorithm pre-allocates resources for time-sensitive periodic streams based on the reported \acp{BD}, ensuring alignment with the \ac{TSN} mechanisms \acl{TAS} and \acl{PSFP}. Meanwhile, remaining resources are dynamically allocated to non-deterministic flows using established strategies such as \acl{PF}, Max C/I, or a \acl{QoS}-aware priority-based scheduler.
The scheduler's performance is evaluated through OMNeT++ simulations. 
The results demonstrate support for diverse \ac{TSN} flows while ensuring deadline-aware scheduling of time-sensitive \ac{UL} traffic in mobility scenarios. Periodic time-sensitive flows are end-to-end scheduled across domains, improving the resource efficiency by 28\% compared to the \acl{CG} baseline. While reliability is preserved, non-deterministic rate-sensitive flows benefit from the improved resource utilization, resulting in higher throughput.

\acresetall
\end{abstract}


\section{Introduction} \label{sec:intro}
The advancement of Industry 4.0 increases the demand for wireless real-time communication in industrial scenarios~\cite {WirelessTSN, SurveyTSNULL}. Hence, communication systems that provide low-latency and deterministic behavior across wireless and wired domains are essential to support applications such as automated guided vehicles, IoT, and robot control. The Ethernet standards of \ac{TSN} address these requirements in wired networks, providing a large set of traffic scheduling and reliability mechanisms to ensure \ac{QoS}-aware communication. 

The \ac{TAS} guarantees a deterministic low-latency communication for periodic traffic in time-aware \ac{TSN} systems~\cite{8021Qbv}. By providing dedicated transmission slots, the \ac{TAS} schedule ensures end-to-end communication with negligible queuing delays for periodic, time-critical flows. To protect the schedule from misbehaving packets, the \ac{TAS} is typically combined with the \ac{PSFP} mechanism~\cite{8021Qci}, which filters packets that violate timing or rate requirements.  

To extend the deterministic low-latency capabilities of \ac{TSN} to wireless networks, integrating 5G as a transparent bridge is a promising approach \cite{WirelessTSN, SurveyTSNULL}. Beyond providing flexibility and mobility, 5G inherently supports \ac{URLLC}. 
In particular, private 5G networks are well-suited for industrial applications, offering customized configurations and operating on a dedicated spectrum, commonly relying on \ac{TDD} mode.

\begin{figure}[!tb]
    \centering
    \includegraphics[width=1.0\linewidth]{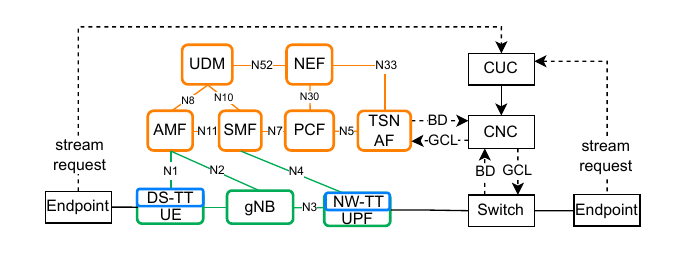}
  %  \vspace{-0.7cm}
    \caption{Integration of 5G as transparent bridge within a fully centralized \ac{TSN} system according to 3GPP  \protect{\cite{23.501}}.}
    \label{fig:architecture}
\end{figure}

3GPP release 16 proposes the integration of 5G with a fully centralized \ac{TSN} system~\cite{23.501,8021Qcc}, as shown in Fig.~\ref{fig:architecture}. The \ac{CNC} module calculates end-to-end schedules such as the \ac{TAS} schedule based on the reported \acp{BD} of all switches and provides bridge configurations. Hence, the pre-calculation of the 5G \ac{BD} is crucial for the transparent integration. 

In contrast to classical \ac{TSN} switches, where the \ac{BD} is mainly affected by higher-priority traffic, the 5G \ac{UL} delay is strongly influenced by the unidirectional \ac{TDD}-based transmission medium and dynamic resource allocation~\cite {Raffeck, Adamuz}. 
Although 3GPP introduced mini-slots to address the limitations of classical \ac{TDD} patterns prioritizing \ac{DL} traffic, most \ac{TDD}-based 5G networks and \ac{COT} products still rely on conventional patterns. While the majority of industrial use cases would benefit from a variety of \ac{TDD} patterns~\cite{NGMN_5G_TDD_Uplink_WhitePaper_2021}, regulatory constraints related to band harmonization and coexistence with 4G currently restrict deployments to using classical \ac{TDD} patterns~\cite{ECC_Report_296, ITU-Report2499}.
Therefore, the pre-calculation of the 5G \ac{BD} must be applicable to any kind of \ac{TDD} pattern. 

To reduce the \ac{UL} delay of time-sensitive flows, 3GPP standardized \acp{CG}, which periodically reserve \acp{RB} for a given \ac{UE}~\cite{38.321,38.331}. \acp{CG} rely on a pre-configured \ac{MCS} to guarantee reserved resources and improve predictability. However, a static \ac{MCS} that does not adapt to channel conditions can lead to over-provisioning or reduced reliability. Additionally, grant-free static scheduling is unsuitable for non-deterministic flows, which mainly require a certain throughput rather than a deadline. 

To fill this gap, this paper proposes a heuristic \ac{UL} scheduler that combines static and dynamic resource allocation to serve the heterogeneous traffic of industrial communication. The scheduler computes the provided 5G \ac{BD} based on the allocation principle and the \ac{TDD} pattern to classify \ac{TSN} \acp{TC} as either dynamically or statically scheduled, assigning the corresponding \ac{BD} accordingly. Using the \ac{PSFP} schedule provided by the \ac{CNC} module, the heuristic scheduler pre-allocates radio resources for periodic streams, ensuring that the reported \acp{BD} are met and aligning the transmissions with the \ac{TAS} schedule. To overcome the limitations of \acp{CG}, we propose a pre-allocation method, which adapts the reserved resources to current channel conditions. 

The scheduler is analyzed through OMNeT++ simulations using the P5G-TSN framework~\cite{p5gtsn}. Key aspects such as \ac{UL} control signaling, \ac{TDD} pattern configuration, and channel conditions are modeled to reflect realistic deployment scenarios.

Specifically, the contributions of this paper are:
\begin{itemize}
    \item For time-aware 5G-\ac{TSN} systems, a heuristic \ac{UL} scheduler is derived by considering only the applied \ac{PSFP} schedule of the \ac{CNC}, preserving the concept of a transparent integration. 
    \item The proposed \ac{UL} scheduler is based on grant-free and grant-based allocations, enabling different \acp{BD} for sporadic and time-sensitive streams.
    \item The grant-free allocation guarantees a low-latency and deterministic \ac{BD} for periodic flows under varying channel conditions for any kind of \ac{TDD} patterns, enabling joint scheduling with the \ac{TAS}.
    %\item The grant-based scheduling discipline can be chosen independently from the grant-free allocations. The performance of different grant-based schedulers is evaluated to allocate resources for sporadic flows. 
    %Hence, different disciplines of state-of-the-art, including Max C/I, \ac{PF}, and a priority-based scheduler, are considered and improved in terms of latency by considering the maximum \ac{BS} to allocate resources of latency sensitive sporadic flows. 
\end{itemize}

The remainder of the paper is organized as follows: Section~\ref{sec:soa} provides an overview of the state-of-the-art on end-to-end scheduling in 5G-\ac{TSN}. Section~\ref{sec:background} introduces the background on 5G integration with time-aware \ac{TSN} and discusses the requirements of industrial applications. Section~\ref{sec:scheduler} presents the proposed heuristic scheduler, including the computation of 5G \ac{BD}. Section~\ref{sec:scenario} describes the simulation scenarios and discusses the results. Finally, Section~\ref{sec:conclusion} concludes the paper.
\section{State-of-the-Art} \label{sec:soa}
This section reviews the state-of-the-art on 5G \ac{BD} estimation, scheduling of \acp{CG}, and joint schedulers in 5G-\ac{TSN}.

\subsection{5G Bridge Delay}
Ambrosy et al.~\cite{Ambrosy_PDB} and Larrañaga et al.~\cite{Larranga_BD} calculate the required \ac{PDB} of a 5G link to define the 5G \ac{BD}. Their calculations consider aspects such as the application's deadline requirements, the wired path delay, and the residence time in the \ac{TT}. However, a discussion on which 5G \ac{BD} can be guaranteed, depending on the \ac{RAN} scheduler, the \ac{TDD} pattern, and the numerology, is missing.

Maghsoudnia et al.~\cite{Maghsoudnia} discuss this aspect, addressing which \ac{TDD} configurations satisfy \ac{URLLC} requirements. Their results show that only mini-slots or grant-free scheduling meet the \ac{UL} delay constraints. Since their focus is on minimizing delay through optimal \ac{TDD} patterns, the calculation of the \ac{BD} for a pre-configured pattern, as in private 5G, remains a missing component, as does the definition of a grant-free scheduler that allocates resources accordingly.

\subsection{Grant-free Scheduling of Time-Sensitive Traffic}
Several studies investigate resource allocations based on \acp{CG} for time-sensitive traffic. Larrañaga et al.~\cite{Larrañaga} propose an iterative scheduler that calculates the resource schedule for a single hyperperiod (the least common multiple of all flows). It reserves multiple grants per stream and iteratively changes the embedding order to find an optimal schedule. However, the algorithm is limited to a static channel and an \ac{TDD} pattern with a single special slot, which makes it unsuitable for mobility scenarios as well as for private 5G deployments that typically rely on conventional \ac{TDD} patterns. 

Tsakos et al.~\cite{Tsakos} extend this concept with a heuristic scheduler that adapts to varying channel conditions by allocating a single \ac{CG} per stream and using proactive monitoring to detect resource conflicts caused by different flow periodicities or delegated channels. While this enhances applicability, it assumes a single-slot \ac{TDD} pattern, which is unsuitable for private 5G, and always schedules flows for immediate transmission, regardless of their specific deadline and jitter requirements.

\subsection{Joint Scheduling of 5G and Time-Aware TSN}
Rodriguez-Martin et al.~\cite{Rodriguez} adapt the \ac{TAS} schedule to the 5G jitter by adjusting the schedule's offset and time slot length. While the method achieves a deterministic end-to-end delay of 20~\si{ms}, it is insufficient for some \ac{TSN} \acp{TC}, underlining the need for low-latency 5G schedulers.

Ginthör et al.~\cite{Ginthör} first propose an optimization problem to jointly schedule \ac{RAN} resources and the \ac{TAS} schedule. The approach considers several 5G-specific characteristics, such as half-duplex constraints and parallel packet transmissions of multiple \acp{UE}, but neglects the \ac{TDD} pattern. Moreover, the optimization requires the \ac{CNC} module to be aware of the 5G link, which breaks the principle of transparent integration. Due to the computational complexity inherent in an optimization problem, it cannot be applied for runtime scheduling, highlighting the demand for a heuristic scheduler.

Zhang et al.~\cite{ZhangYang} introduce a deep reinforcement learning approach that adapts the \ac{DL} resource scheduling  to current traffic demand and the \ac{TAS} schedule. Although the approach is promising, it focuses only on \ac{DL}, leaving the impact of the \ac{UL} characteristics and control signaling unverified.

The authors of~\cite{Zhang} propose a framework featuring a multi-priority scheduler that pre-allocates radio resources according to the \ac{TAS} schedule while dynamically allocating resources for best-effort flows.
A \ac{PSFP}-based injection mechanism adjusts frame priorities after 5G transmissions to compensate for jitter.
Although the approach achieves a low end-to-end delay, it does not consider the TDD pattern, configurable numerologies, or a channel model, which may limit its general applicability.

Debnath et al.~\cite{Debnath} propose an \ac{QoS} algorithm to map \ac{TSN} streams on standardized 5G profiles and evaluate the resulting mapping exemplarily through OMNeT++ simulations. The dynamic resource scheduler applies the priority, or the \ac{PDB} of each 5G profile, in a strict-priority manner. Although the study provides detailed simulation results, the evaluation is limited to \ac{DL} traffic and single-slot \ac{TDD} patterns, making it unsuitable for private 5G deployments.

Although extensive research has been conducted on joint resource allocation in 5G-\ac{TSN}, existing solutions  fail to account for varying channel conditions, neglect the impact of the \ac{TDD} pattern, or compromise transparent integration. To address these limitations, we propose a joint resource allocation approach based on the \ac{TAS} and a heuristic heterogeneous \ac{RAN} scheduler that combines grant-free and grant-based scheduling while explicitly considering the \ac{TDD} pattern. The applied pre-allocation mechanism ensures a deterministic 5G \ac{BD}, enabling joint scheduling with the \ac{TAS}. To maintain transparent integration, resource allocation relies solely on information from the \ac{PSFP} schedule, as the \ac{CNC} configures the 5G link in the same manner as any other \ac{TSN} bridge. The scheduler’s heuristic design and awareness of the \ac{TDD} pattern make it suitable for evaluation in practical testbeds.
\section{Background} \label{sec:background}
In this section, we introduce mechanisms of time-aware \ac{TSN} and describe the architectural integration of 5G as a transparent bridge. We then discuss 5G radio resource allocation, distinguishing between grant-based and grant-free schemes. Finally, we provide an overview of classical industrial applications, highlighting typical requirements per \ac{TC}.
\subsection{Time-Aware TSN}
\begin{figure}[!tb]
    \centering
    \includegraphics[width=1.0\linewidth]{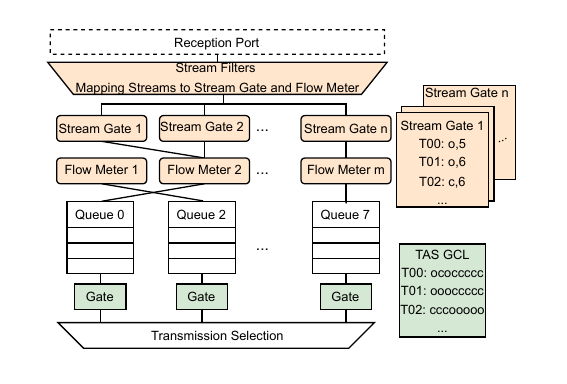}
    \vspace{-0.7cm}
    \caption{Workflow of time-aware \ac{TSN} using \ac{PSFP} to protect the \ac{TAS} schedule (abstracted from~\protect{\cite{8021Qbv}}, \protect{\cite{8021Qci}}).}
    \label{fig:PSFP+TAS}
\end{figure}

The \ac{TAS}, firstly introduced in IEEE 802.1 Qbv~\cite{8021Qbv}, provides a time-slotted access to the transmission medium.
As shown in Fig.~\ref{fig:PSFP+TAS}, the \ac{GCL} defines which \acp{TC} are allowed to transmit during each slot. This mechanism is particularly beneficial for time-sensitive traffic, as it enables end-to-end scheduling by reserving dedicated transmission slots along the communication path. 

However, in case multiple flows share the same \ac{TC}, a misbehaving stream may disrupt the \ac{TAS} schedule by transmitting outside its dedicated slots. Hence, the \ac{TAS} schedule is typically protected  by the reliability mechanism \ac{PSFP}, specified in IEEE 802.1Qci~\cite{8021Qci}. As shown in Fig.~\ref{fig:PSFP+TAS}, \ac{PSFP} operates as an incoming filter. Stream gates drop frames according to the \ac{GCL}, defining in which time intervals frames are allowed to pass and optionally assigning an Internal Priority Value. This priority value overrules the priority in the 802.1Q header and determines the selection of the queue. 

The flow meters regulate traffic based on stream rate and \ac{BS} using a two-rate, three-color marking token bucket~\cite{8021Qci, MEF10.3}. The green bucket increases its tokens according to the \ac{CIR} up to the \ac{CBS}. Optionally, a yellow bucket can be configured. If the stream exceeds its configured rate, the packet is marked red and dropped. 
% is ptp discussion here required?

\subsection{Integration of 5G with TSN}
5G is integrated as a transparent bridge within a fully centralized \ac{TSN} system (see Fig.~\ref{fig:architecture})~\cite{8021Qcc, 23.501}. In this centralized configuration, end stations send their formatted stream requirements to the \ac{CUC} module. These requirements include parameters such as the sending interval and \ac{BS}, as well as optional time-aware information, including the transmission offset and maximum jitter. 

The \ac{CNC} module aggregates bridge capabilities, including minimum and maximum \acp{BD} defined per port pair and \ac{TC}, distinguishing between delays that are dependent and independent of the frame size~\cite{8021Qcc}. Based on stream requirements provided by the \ac{CUC} and bridge capabilities, the \ac{CNC} calculates the schedules and configurations across the network. 

Within the 5G-\ac{TSN} architecture, the \ac{DS-TT} and \ac{NW-TT} handle the translation between the user plane and \ac{TSN} domain, while in the control plane, the \ac{TSN} \ac{AF} reports the 5G bridge capabilities~\cite{23.501}.
Due to the principle of transparent integration, the \ac{CNC} does not provide 5G-specific scheduling information. According to 3GPP specifications~\cite{23.501}, this limitation is resolved by mapping \ac{TSN} streams to 5G \ac{QoS} profiles. Each 5G profile is identified by a 5QI value and specifies parameters such as resource type, priority, \ac{PDB}, \ac{PER}, and \ac{MDBV}. While several 5G profiles are standardized, additional profiles can be defined to support specific industrial requirements. 

 The \ac{QoS} mapping is not standardized and has therefore been explored in related research. Satka et al. \cite{Satka} and Debnath et al. \cite{Debnath} propose algorithms that map \ac{TSN} streams to standardized profiles, whereas Ambrosy et al.~\cite{Ambrosy_QoS} define additional 5G profiles to better represent \ac{TSN} traffic types. 
 
Although this mapping allows the translation of \ac{QoS} parameters, it does not ensure the end-to-end requirements or a deterministic 5G \ac{BD}. The radio resource scheduler remains responsible for allocating resources according to the profile requirements, which is beyond the scope of standardization. 

Optionally, the \acp{TT} can support the \ac{PSFP} mechanism, providing ingress filtering, which is reported as additional bridge capability by the \ac{TSN} \ac{AF}~\cite{23.501}. The calculated \ac{PSFP} \ac{GCL} of the \ac{CNC} reveals traffic parameters such as \ac{BAT}, \ac{BS}, and periodicity T. For \ac{DC-GBR} flows, these parameters can be stored in a \ac{TSCAI} and shared with the 5G \ac{RAN} scheduler. In the following, flows that are considered as \ac{DC-GBR} are described by the following tuple: $f_i = \{BAT_i, BS_i, T_i, TC_i\}$. 
%3GPP proposes the usage of Hold-and-Forwarding buffering in \cite{23.501} to compensate the jitter of the 5G bridge. The concept is based on the same as the \ac{TAS}, providing sequential dedicated transmission slots to \acp{TC}. To avoid packets missing their dedicated slots, the Hold-and-Forwarding buffering has to consider the worst case delay of the 5G system to provide afterwards a deterministic delay. 

\subsection{5G Radio Resource Allocation}
In 5G \acp{RAN}, the radio resource allocation is significantly impacted by the \ac{TDD} pattern, which defines the available transmission opportunities for data and control signaling in time~\cite{Raffeck,Adamuz}. A \ac{TDD} pattern consists of $s_{dl}$ \ac{DL} and $s_{ul}$ \ac{UL} slots separated by $s_{sp}$ special slots ensuring interference-free switching between the transmission directions~\cite{38.211}. 
The duration of a slot depends on the numerology $\mu$ and is defined as $T_{slot} = 2^{-\mu}$, which in turn determines the time duration of a \ac{TDD} pattern $T_{tdd}$. 
In this paper a \ac{TDD} pattern is described using the tuple $\{s_{dl}, s_{sp}, s_{ul}, T_{slot}, T_{tdd}\}$.

The \ac{gNB} distributes the \acp{RB} of a slot among the backlogged \acp{UE}, which are then mapped to an effective \ac{TBS} based on the measured channel quality~\cite{38.214}.
For \ac{UL} scheduling, the \ac{gNB} determines the \ac{CQI} from the received transmission power. Each \ac{CQI} value is mapped to a standardized \ac{MCS}, which defines the modulation order and code rate required to meet the target Block Error Rate (BLER). 
The modulation order defines the number of bits per \ac{RE}, while the code rate represents the ratio of useful to total transmitted bits, controlling the redundancy of each transmission. The \ac{TBS} is calculated following~\cite{38.214}, based on the selected \ac{MCS} and the number of allocated \acp{RB}.

To serve many \acp{UE} and optimize the bandwidth utilization, grant-based dynamic resource allocation is typically deployed. To request resources, the \ac{UE} attaches the \ac{BSR} to its next transmission or, if no valid grant is available, it sends a \ac{SR}~\cite{38.321}. Consequently, the \ac{gNB} adds the \ac{UE} to the list of backlogged \acp{UE} that are considered in the next scheduling decision.

When the \ac{gNB} receives a \ac{BSR}, it allocates \acp{RB} based on the buffered data~\cite{38.321}. If a \ac{SR} is received, the \ac{gNB} allocates an initial grant with a pre-configured number of \acp{RB}. Using this initial grant, the \ac{UE} transmits a part of its buffered data along with a \ac{BSR}, such that the \ac{gNB} allocates appropriate resources in the subsequent \ac{TDD} cycle. Because it consumes only a small amount of \acp{RB} while allowing the transmission of a \ac{BSR}, the scheduler is typically implemented to always allocate the initial grant, regardless of the scheduling discipline.

To support time-sensitive periodic traffic, 5G also provides \acp{CG}, which periodically reserve radio resources~\cite{38.321,38.331}. Each \ac{CG} specifies the frequency resources allocated within a slot for a fixed \ac{MCS}. Consequently, the \ac{MCS} and the number of \acp{RB} are not dynamically adapted to the current \ac{CQI}, which contrasts with dynamic scheduling. 

\subsection{Industrial Use Cases}

\begin{table*}[!tb]
\caption{An overview of \textbf{p}eriodic and \textbf{s}poradic traffic types discussed in IEC/IEEE 60802~\protect{\cite{acia}}.}
\centering
\small
\begin{tabular}{|l|c|c|c|c|c|c|c|c|c|c|c|c|c|}
\hline
\textbf{TC} & \textbf{Traffic Type} & \textbf{P/S} & \textbf{Period}& \textbf{Size (B)}  & \textbf{Guarantees} & \textbf{TSN Shaper \& Scheduler} \\
\hline
7 & Network Control          & p & 50\si{ms}--1\si{s}      & variable: 50--500  & throughput & \\
\hline
6 & Isochronous           & p & 100\si{\mu s}--2\si{ms}& fixed: 30--100 & jitter free \& deadline & TAS \& time-based PSFP  \\
\hline
5 & Cyclic-Sync. \& Async.  & p & 500\si{\mu s}--20\si{ms}&  fixed: 50--1000  & bounded jitter \& latency & TAS \& time-based PSFP \\
\hline
4 & Control Events        & s & 10\si{ms}--50\si{ms}   & variable: 100--200 & bounded latency & rate-based PSFP \\
\hline
3 & Alarms \& Commands       &s & 2\si{s}             & variable: 100--1500 & bounded latency   & rate-based PSFP \\
\hline
2 & Config \& Diagnostics             & s & n.a.            & variable: 500--1500 & throughput  & rate-based PSFP \\
\hline
1 & Audio \& Video        & p &  rate    & variable: 1000--1500 & throughput  & rate-based PSFP \\
\hline
0 & Best Effort           & s & n.a.    & variable: 30--1500  & &\\
\hline
\end{tabular}
\label{tab:traffic_types}
\end{table*}
The ongoing project IEC/IEEE 60802 specifies traffic types for industrial automation applications. Table~\ref{tab:traffic_types} provides an overview of these traffic types, which are grouped into \acp{TC}, following the definition of 5G-ACIA~\cite{acia}. \acp{TC} with a periodic sending interval are further classified based on \ac{QoS} requirements: jitter-free delivery within deadlines, bounded jitter and delay, or guaranteed throughput. Non-deterministic \acp{TC} (sporadic sending or variable packet size) partially require either a throughput or a bounded latency. 

To satisfy the requirements of time-sensitive periodic \acp{TC} in the wired domain, 5G-ACIA recommends using the \ac{TAS} with a time-based \ac{PSFP}. Due to its time-slotted, cyclic behavior, the \ac{TAS} applies only to periodic \acp{TC}, whereas non-deterministic flows are controlled by the rate-based \ac{PSFP} mechanism.
\section{Joint Resource Allocation in 5G-TSN Systems} \label{sec:scheduler}
The integration of 5G within a time-aware \ac{TSN} system significantly depends on the calculation of the 5G \ac{BD} per \ac{TC}. As shown in the literature review, previous studies have computed the required 5G \acp{BD} depending on the remaining path delay and the end-to-end requirements~\cite{Larranga_BD, Ambrosy_PDB}. However, the calculation of the \ac{UL} \acp{BD} with respect to the \ac{TDD} pattern and the design of a scheduler that allocates resources accordingly remain open challenges. 

We propose a heterogeneous scheduler in the form of a heuristic scheduler consisting of two parts: (1) it pre-allocates radio resources for \acp{TC} requiring static allocations and (2) distributes residual resources dynamically among flows of non-deterministic \acp{TC}. In the following, we first describe the calculation of the 5G \acp{BD}, then propose a heuristic algorithm to pre-allocate radio resources, and finally discuss the scheduling of residual resources for non-deterministic \acp{TC}.

\subsection{5G Bridge Delay}
\begin{figure}[!b]
    \centering
    \includegraphics[width=1.0\linewidth]{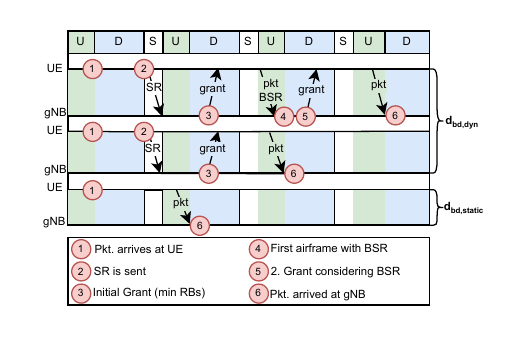}
   % \vspace{-1cm}
    \caption{Delay of dynamic and static resource allocation in \ac{TDD}-based 5G.}
    \label{fig:UL_Scheduling}
\end{figure}

The selection of the 5G \acp{BD} directly affects the end-to-end \ac{TAS} schedule. If the \ac{BD} is too small, packets may miss their dedicated slots, but if the \ac{BD} is too large, the end-to-end delay increases unnecessarily. 
Following \cite{Larranga_BD}, the 5G \ac{BD} is described by~\eqref{Eq:BD_general}. It comprises the forwarding delay between \ac{TT} and \ac{UE} $\delta$, as well as the delay that is dependent and independent of the frame size.
\begin{equation} \label{Eq:BD_general}
    d_{bd} = \delta + d_{dependent} + d_{independent}
\end{equation} 
Moreover, we also need to account for the influence of the \ac{TDD} pattern and the type of resource allocation in \ac{BD} calculation. These details are illustrated in Fig.~\ref{fig:UL_Scheduling}, distinguishing between static and dynamic resource allocation.

In dynamic scheduling, the maximum \ac{BD} occurs when a packet arrives during the last \ac{UL} slot of the \ac{TDD} cycle, such that the \ac{UE} must wait for the subsequent cycle to send the \ac{SR}. In the worst case, the initial grant is insufficient to transmit the entire packet, requiring an additional grant. Consequently, the maximum \ac{BD} includes one slot for buffering the packet at the \ac{UE}, one \ac{TDD} cycle to request the grant, and two cycles to complete the transmission, as calculated in~\eqref{Eq:dyn_max}. 

If the initial grant is sufficient to transmit the entire packet, the \ac{BD} is reduced by the length of one \ac{TDD} cycle. Since the \ac{SR} does not contain buffer information, this reduction can only be guaranteed if the \ac{gNB} knows the \ac{BS} in advance. Nevertheless, the described maximum \acp{BD} can be only achieved under the assumption of no retransmissions and that the scheduler schedules the \ac{UE} in the subsequent \ac{TDD} cycles. 
\begin{equation}\label{Eq:dyn_max}
max\{d_{bd, dyn}\}= \\
\begin{cases}
    \delta + 3 T_{tdd} + T_{slot} & \text{default}\\
    \delta + 2 T_{tdd} + T_{slot}  & \text{BS known}
\end{cases}
\end{equation}

The minimal \ac{BD} occurs when the frame arrives at the \ac{UE} immediately before the last \ac{UL} slot of the \ac{TDD} cycle, allowing the \ac{UE} to send the  \ac{SR} without any delay. In addition, the \ac{gNB} has to assign the required resources in the first available special slot to ensure the minimal \ac{BD} defined by~\eqref{Eq:dyn_min}.
\begin{equation} \label{Eq:dyn_min}
min\{d_{bd, dyn}\}= \\
\begin{cases}
    \delta + T_{slot} (s_{dl}+2) + T_{tdd} & \text{default}\\
    \delta + T_{slot} (s_{dl}+2)  & \text{BS known}
\end{cases}
\end{equation}

The \ac{BD} of static grant-free scheduling depends on the arrival time of  frames at the \ac{UE} relative to the \ac{TDD} pattern. Since no resources have to be requested, the \ac{UE} only waits for its pre-allocated \ac{UL} slot. Under the assumption that the \ac{gNB} pre-allocates resources for the \ac{UE} in the next \ac{TDD} cycle, the corresponding maximum \ac{BD} can be derived as shown in~\eqref{Eq:static_max}. 
\begin{equation} \label{Eq:static_max}
    max\{d_{bd,static}\} = \delta + T_{tdd} + T_{slot}
\end{equation}

If the packet arrives immediately before the pre-allocated slot, the \ac{BD} is minimized to the sum of one slot and the forwarding delay. 
\begin{equation} \label{Eq:static_min}
    min\{d_{bd,static}\} = \delta + T_{slot}
\end{equation}

To guarantee this minimum \ac{BD} for all packets of a periodic stream $f_i$, assuming sufficient radio resources are pre-allocated in the slot $x$, the following preconditions have to be fulfilled by the $BAT_i$ and the periodicity $T_i$:
\begin{equation} \label{Eq:Cond1}
 BAT_i \mod T_{tdd} = x  T_{slot}-\delta \\
\end{equation}
\begin{equation}
  T_i \mod T_{tdd} = 0
\label{Eq:Cond2}
\end{equation}

To determine the appropriate allocation method and to assign the corresponding \ac{BD} for each \ac{TC}, the traffic characteristics as defined by 5G-ACIA (Table~\ref{tab:traffic_types}) are taken into account. Meanwhile, Table~\ref{tab:BDs} summarizes the resulting mapping and the information required by the resource scheduler. 

Due to their deterministic sending behavior in terms of sending interval and packet size, along with their bounded latency and jitter requirements, \ac{TC} 5 and 6 are scheduled grant-free. To achieve a jitter-free communication for \ac{TC} 6, the maximum and minimum \ac{BD} must be identical. Instead of the de-jittering approach of Hold-and-Forwarding buffering~\cite{23.501}, we propose achieving jitter-free communication by scheduling transmissions always in the same slot. While Hold-and-Forwarding buffering would delay all frames to the maximum delay of static allocation, our approach ensures not only jitter-free but also deadline-aware transmissions. To reduce the maximum \ac{BD} of \ac{TC} 6 to \eqref{Eq:static_min}, the conditions defined by \eqref{Eq:Cond1} and \eqref{Eq:Cond2} have to be fulfilled. If a stream fails to meet either condition, the 5G \ac{TSN} \ac{AF} reports an increased \ac{BD} to the \ac{CNC}, triggering a reconfiguration. 
%Hence, streams of \ac{TC} 6 have to fulfill the condition given in~\eqref{Eq:Cond2}. 
%Although this introduces additional control overhead, it remains compliant with the fully centralized configuration module of \ac{TSN} \cite{8021Q} and preserves the concept of a transparent 5G bridge. \\

For \acp{TC} with a non-deterministic sending interval or variable packet size, static allocation is inefficient, making dynamic resource allocation the preferred approach. To reduce the delay of time-sensitive flows with non-deterministic sending behavior, the scheduler must know the \ac{BS} of a flow in advance to adapt the initial grant accordingly. Although the traffic patterns are only communicated to the \ac{CUC}, the 5G bridge can still determine the maximum \ac{BS} by taking the rate-based \ac{PSFP} schedule into account, which limits the \ac{BS} and rate at the \acp{TT}. To convey this information to the resource scheduler, the \ac{MDBV} parameter of the configured 5G profile can be updated. To reduce the risk of resource over-provisioning, this method is applied only to \ac{TC}~4, which features a narrow packet size distribution, as shown in Table~\ref{tab:traffic_types}.

\begin{table}[!tb]
\caption{5G \acp{BD} per \ac{TC} and its preconditions in terms of traffic patterns and required information at the \ac{RAN} scheduler.}
\begin{tabular}{|l|c|c|c|c|}
\hline
\textbf{TCs} & \multicolumn{2}{c|}{\textbf{5G BD}} & \multicolumn{2}{c|}{\textbf{Conditions to fulfill 5G BD}} \\
\hline
 & \textbf{min$\{d_{bd}\}$} & \textbf{max$\{d_{bd}\}$}
& \textbf{Traffic} & \textbf{5G Information} \\
\hline
6 &  \multicolumn{2}{c|}{\eqref{Eq:static_min}} & ~\eqref{Eq:Cond1}, \eqref{Eq:Cond2} & \ac{TSCAI} \\
\hline
5 & \eqref{Eq:static_min} & \eqref{Eq:static_max} & periodic $T_i$ & \ac{TSCAI} \\
\hline
4 & \eqref{Eq:dyn_min} & \eqref{Eq:dyn_max} & & \ac{MDBV}\\
\hline
0-3,7&  \eqref{Eq:dyn_min} & \eqref{Eq:dyn_max} & & \\
\hline
\end{tabular}
\label{tab:BDs}
\end{table}

\subsection{Static Grant-Free Resource Allocation}
The static allocation component of our heterogeneous scheduler pre-allocates radio resources for $N_F$ periodic flows according to reported \acp{BD}. Algorithm~\ref{alg:cg} details the pre-allocation, which relies on \ac{EDF} scheduling, while respecting the \ac{TDD} pattern constraints. 

Inspired by~\cite{Larrañaga}, the schedule is computed per hyperperiod (line 1). After initializing the resource grid, Algorithm~\ref{alg:reset} determines each flow’s arrival time and total bytes to be scheduled. During hyperperiod iteration (line 6), a scheduling window $w$ is defined for newly arriving flows, specifying the time restrictions to schedule a flow (line 10).

Based on the scheduling window $w$, the flows are sorted in order of earliest deadline first (line 13). If flows share the same window size, flows with harder restrictions in terms of radio resource availability are prioritized. In this paper, we assume that the \ac{CQI} is measured per \ac{RB}, allowing per-\ac{RB} scheduling decisions. To calculate the effective \ac{MCS} for a transmission, the \ac{gNB} computes the mean \ac{CQI} across all \acp{RB}. Only \acp{RB} with a \ac{CQI} equal to or higher than the mean are considered for allocation, ensuring more reliable transmissions. Hence, in line 13 of Algorithm~\ref{alg:cg}, \acp{UE} with fewer available \acp{RB} are prioritized to ensure that all flows can be scheduled.
In line~16, \acp{RB} are allocated for transmission based on the available \acp{RB}, the required \ac{BS}, and the \ac{CQI}. The \ac{CQI} is mapped to an \ac{MCS}, which might be restricted in range. If resources can be scheduled for the \ac{UE}, the function returns the amount of scheduled RBs and the served payload by the \ac{TBS}.

Following the centralized \ac{TSN} configuration model, the 5G bridge must first verify whether all configured flows can be supported, even under worst-case conditions. To this end, a minimum \ac{MCS} is defined, and the system ensures that all flows remain schedulable even if every \ac{UE} operates at this minimum \ac{MCS}. Although this approach limits the number of scheduled time-sensitive flows, it ensures reliable transmissions. Moreover, in typical industrial scenarios where non-deterministic flows are also present, the residual resources can still be effectively utilized. 

Following the concept of \acp{CG}, Algorithm~\ref{alg:cg} is executed once during the network planning phase, restricting the \ac{MCS} to a single value in line 16. In contrast, we propose an adaptive \ac{MCS} pre-allocation scheme that executes Algorithm~\ref{alg:cg} during network operation to dynamically update the grants only when the schedule must be adjusted due to degraded channel conditions. Although this approach introduces additional control overhead compared to static \acp{CG}, it enhances reliability and throughput.

\begin{algorithm}[!tb]
    \caption{Pre-allocation of radio resource for periodic time-sensitive flows.}\label{alg:cg}
\begin{algorithmic}[1]
\STATE{$HP = lcm(T_{tdd}, T_1, T_2, ..., T_{NF})$}
\STATE{ $RB[i][j] = \#PRB$ for all i in $[0,HP]$}
\FOR{$f_i \in F$}
    \STATE{$ResetFlow(f_i)$}
\ENDFOR
\FOR{$slot\in [0,HP]$}
    \IF{$slot$ is UL or special}
    \FOR{$f_i \in F$}
        \IF{$f_i$ arrived $\land~w_i == -1$}
            \STATE{$w_i = max\{d_{bd}\}$}
        \ENDIF
    \ENDFOR
    \STATE{ $F  \gets SortByConstrainedness(F)$}
    \FOR{$f_i \in F$}
        \IF{$BS_i^{req} > 0 \land w_i > 0$}
            \STATE{$RB[i][j],BS_i^{req}\mathrel{-}= \text{Calc\_RBs}(RB[i][j], $}\\
\hspace{0.5cm} $BS_i^{req}, Cqi,\text{MCS}_{min}, \text{MCS}_{max})$
        \ENDIF
    \STATE{$w_i\mathrel{--}$}
    \IF{$w_i <1 \lor BS_i^{req} \leq 0$}
        \STATE{$BAT_i \mathrel{+}= T_i$}
        \STATE{$ResetFlow(f_i)$}
    \ENDIF
\ENDFOR
\ENDIF
\ENDFOR
\end{algorithmic}
\label{algo}
\end{algorithm}
\begin{algorithm}[!tb]
\caption{Function $ResetFlow(f_i)$ calculates the arrival time point of the flow with respect to the \ac{TDD} pattern.}\label{alg:reset}
\begin{algorithmic}[1]
   \STATE{$ BAT_i^{frame} = \Big\lfloor\frac{BAT_i}{T_{tdd}}\Big\rfloor$}
   \STATE{$BAT_i^{slot} = \Big\lceil BAT_i - \frac{BAT_i^{frame} \cdot T_{slot}}{T_{tdd}}\Big\rceil $}
   \STATE{$BS_i^{req} = BS_i$, $w_i = -1$}
\end{algorithmic}
\end{algorithm}

\subsection{Dynamic Grant-Based Resource Allocation}
To support diverse flows, our proposed heterogeneous scheduler pre-allocates resources for time-sensitive,  deterministic flows and dynamically allocates the residual resources to non-deterministic flows on demand. Consequently, unreserved resources can be managed by any state-of-the-art grant-based scheduler, allowing for the selection of a dynamic scheduling discipline tailored to the system's requirements. Furthermore, dynamically scheduled flows benefit from the improved resource utilization provided by our adaptive \ac{MCS} pre-allocation scheme compared with \acp{CG}.
%In the wired domain, non-deterministic flows are typically managed using strict-priority scheduling. Translating this concept to the 5G domain necessitates a dynamic, priority-based scheduler that accounts for the priority levels of 5G profiles, as proposed in~\cite{Debnath}. However, the channel conditions of each \ac{UE} vary depending on its position and transmission power, leading to different effective data rates. Therefore, we evaluate the performance of a priority-based scheduler against channel-aware scheduling strategies such as Max-CI and \ac{PF}. 

Configuring the \ac{MDBV} parameter of the 5G profiles for time-sensitive, non-deterministic flows exposes the \ac{BS} to the \ac{RAN} scheduler, enabling the scheduler to adjust initial grants to meet the flows' demands  and thereby to reduce the delay. From a workflow perspective, our heterogeneous scheduler first reserves resources for periodic streams through static allocation. Initial grants are then allocated to the \acp{UE} independent of the chosen scheduling strategy, before the remaining resources are distributed according to the selected scheduling principle. 
\section{Performance Evaluation} \label{sec:scenario}
We evaluate the performance of our proposed scheduler through OMNeT++ simulations, which incorporate an extension of the P5G-TSN framework~\cite{p5gtsn}. This framework combines Simu5G~\cite{simu5g} to simulate the 5G user plane, and INET~\cite{inet} to support \ac{TSN} mechanisms. Table~\ref{tab:sim} provides an overview of the two simulation scenarios, referred to as periodic and heterogeneous. 
\begin{table}[!tb]
\caption{Simulation parameters of the periodic and heterogeneous simulation scenarios.}
\centering
\begin{tabular}{|l|c|c|}
\hline
\textbf{Parameters} &\multicolumn{2}{|c|}{\textbf{Simulation Scenarios}} \\
\hline
 & Periodic & Heterogenous \\
\hline
\hline
\#Streams & 28 (TC 5,6)& 48 (TC 0,1,4-7) \\  % simple row, not merged
\hline
Topology & \multicolumn{2}{c|}{Talker - SW1 - UEs - gNB - SW2 - Listener} \\
\hline
TSN scheduler & Strict Priority & Strict Priority + TAS \\
\hline
 &  & GF adaptive MCS +  \\
5G &  1) GF adaptive MCS &  1) Dyn. Strict Priority \\
Scheduler &  2) GF static MCS & 2) Dyn. MaxCI \\
 &  3) Dyn. PDB & 3) Dyn. PF\\
\hline
%\multirowcell{2}[-0.2em][l]{Min. MCS\\for GF allocations}
%  & \multicolumn{2}{c|}{\multirow{2}{*}{\centering 5}} \\
%  & \multicolumn{2}{c|}{} \\ \hline
TDD pattern & \multicolumn{2}{c|}{DDDDDDDSUU, $\mu$ = 1} \\
\hline
\#RBs & 106 & 51 \\  % simple row, not merged
\hline
Data rate ethernet & \multicolumn{2}{c|}{1\si{Gbps}} \\
\hline
Channel model & \multicolumn{2}{c|}{INDOOR\_FACTORY\_DL, Fading: Jakes} \\
\hline
Mobility & \multicolumn{2}{c|}{Random Walk, Speed: 5mps} \\
\hline
Area & \multicolumn{2}{c|}{300m x 150m, gNB Position (1m,1m) } \\
\hline
TxPower & \multicolumn{2}{c|}{24 dBm} \\
\hline
QoS  & \multicolumn{2}{c|}{Non-standardized profiles adapted to} \\
mapping & \multicolumn{2}{c|}{ the traffic types following~\cite{Ambrosy_QoS}}\\
\hline
Simulation runs & \multicolumn{2}{c|}{10 runs, 60s in simulation time per run} \\
\hline
%MCS Table & \multicolumn{2}{c|}{1} \\
%\hline
\end{tabular}
\label{tab:sim}
\end{table} 

\begin{figure}[!b]
    \centering
    % --- First plot ---
     \begin{subfigure}{0.5\textwidth}
      %  \centering
  % \resizebox{\textwidth}{!}{\input{figures/periodic_delay}}
  % This file was created with tikzplotlib v0.10.1.
\begin{tikzpicture}					   
									
\definecolor{green}{RGB}{0,128,1}

\begin{axis}[
height=5cm,
width=5.8cm,
legend cell align={left},
legend columns = 3,
legend style={fill opacity=0.8, draw opacity=1, text opacity=1, draw=gray, legend columns=2, column sep=4pt, legend cell align=left, at={(1.15,1.3)},  font=\footnotesize,},
log basis y={10},
trim axis left,
tick align=outside,
tick pos=left,
x grid style={gray},
xmajorgrids,
xmin=-0.86, xmax=8.8,
xtick style={color=black,font=\footnotesize,},
xtick={2.4,6.9},
xticklabels={TC 6, TC 5},
y grid style={gray},
ylabel={End-to-End Delay [ms]},
ymajorgrids,
ymin=0.5, ymax=400,
ymode=log,
ytick style={color=black,font=\footnotesize,},
xticklabel style = {font=\footnotesize},
yticklabel style  = {font=\footnotesize},
xlabel style = {font=\footnotesize},
ylabel style  = {font=\footnotesize},
ytick={0.01,0.1,1,10,100,1000,10000},
yticklabels={
  \(\displaystyle {10^{-2}}\),
  \(\displaystyle {10^{-1}}\),
  \(\displaystyle {10^{0}}\),
  \(\displaystyle {10^{1}}\),
  \(\displaystyle {10^{2}}\),
  \(\displaystyle {10^{3}}\),
  \(\displaystyle {10^{4}}\)
}
]	  
\addlegendimage{draw=black,ybar,ybar legend,draw=black,fill=white,postaction={pattern=north east lines, pattern color = orange}}
\addlegendentry{GF, adaptive MCS}		 
\addlegendimage{draw=black,ybar,ybar legend,fill=white,postaction={pattern=north west lines, pattern color =  blue}}
\addlegendentry{GF, MCS: 5}
\addlegendimage{draw=black,ybar,ybar legend,fill=white,postaction={pattern=crosshatch, pattern color = green}}	
\addlegendentry{GF, MCS: 11}
\addlegendimage{draw=black,ybar,ybar legend,fill=white,postaction={pattern=north east lines, pattern color = red}}
\addlegendentry{GF, MCS: 20}
\addlegendimage{draw=black,ybar,ybar legend,fill=white,postaction={pattern=north west lines, pattern color = gray}}
\addlegendentry{Dyn. \ac{PDB}}

\addlegendimage{red, thick, dashed,legend image post style={xscale=0.5}}
\addlegendentry{max. path delay}

\path [draw=orange, fill=white, postaction={pattern=north east lines, pattern color=orange}]
(axis cs:0.7,0.607)
--(axis cs:1.3,0.607)
--(axis cs:1.3,0.612)
--(axis cs:0.7,0.612)
--(axis cs:0.7,0.607)
--cycle;
\addplot [black, forget plot]
table {%
1 0.607
1 0.605
};
\addplot [black, forget plot]
table {%
1 0.612
1 0.619
};
\addplot [black, forget plot]
table {%
0.85 0.605
1.15 0.605
};
\addplot [black, forget plot]
table {%
0.85 0.619
1.15 0.619
};
\addplot [black, opacity=1, mark=asterisk, mark size=4, mark options={solid}, only marks, forget plot]
table {%
1 0.639
};
\path [draw=blue, fill=white, postaction={pattern=north west lines, pattern color=blue}]
(axis cs:1.4,0.609)
--(axis cs:2,0.609)
--(axis cs:2,0.617)
--(axis cs:1.4,0.617)
--(axis cs:1.4,0.609)
--cycle;
\addplot [black, forget plot]
table {%
1.7 0.609
1.7 0.605
};
\addplot [black, forget plot]
table {%
1.7 0.617
1.7 0.629
};
\addplot [black, forget plot]
table {%
1.55 0.605
1.85 0.605
};
\addplot [black, forget plot]
table {%
1.55 0.629
1.85 0.629
};
\addplot [black, opacity=1, mark=asterisk, mark size=4, mark options={solid}, only marks, forget plot]
table {%
1.7 0.63
};
\path [draw=green, fill=white, postaction={pattern=crosshatch, pattern color=green}]
(axis cs:2.1,0.608)
--(axis cs:2.7,0.608)
--(axis cs:2.7,0.613)
--(axis cs:2.1,0.613)
--(axis cs:2.1,0.608)
--cycle;
\addplot [black, forget plot]
table {%
2.4 0.608
2.4 0.605
};
\addplot [black, forget plot]
table {%
2.4 0.613
2.4 0.62
};
\addplot [black, forget plot]
table {%
2.25 0.605
2.55 0.605
};
\addplot [black, forget plot]
table {%
2.25 0.62
2.55 0.62
};
\addplot [black, opacity=1, mark=asterisk, mark size=4, mark options={solid}, only marks, forget plot]
table {%
2.4 0.639
};
\path [draw=red, fill=white, postaction={pattern=north east lines, pattern color=red}]
(axis cs:2.8,0.607)
--(axis cs:3.4,0.607)
--(axis cs:3.4,0.611)
--(axis cs:2.8,0.611)
--(axis cs:2.8,0.607)
--cycle;
\addplot [black, forget plot]
table {%
3.1 0.607
3.1 0.605
};
\addplot [black, forget plot]
table {%
3.1 0.611
3.1 0.617
};
\addplot [black, forget plot]
table {%
2.95 0.605
3.25 0.605
};
\addplot [black, forget plot]
table {%
2.95 0.617
3.25 0.617
};
\addplot [black, opacity=1, mark=asterisk, mark size=4, mark options={solid}, only marks, forget plot]
table {%
3.1 0.639
};
\path [draw=gray, fill=white, postaction={pattern=north west lines, pattern color=gray}]
(axis cs:3.5,4.608)
--(axis cs:4.1,4.608)
--(axis cs:4.1,5.112)
--(axis cs:3.5,5.112)
--(axis cs:3.5,4.608)
--cycle;
\addplot [black, forget plot]
table {%
3.8 4.608
3.8 4.605
};
\addplot [black, forget plot]
table {%
3.8 5.112
3.8 5.628
};
\addplot [black, forget plot]
table {%
3.65 4.605
3.95 4.605
};
\addplot [black, forget plot]
table {%
3.65 5.628
3.95 5.628
};
\addplot [black, opacity=1, mark=asterisk, mark size=4, mark options={solid}, only marks, forget plot]
table {%
3.8 85.105
};
\path [draw=orange, fill=white, postaction={pattern=north east lines, pattern color=orange}]
(axis cs:5.2,0.614)
--(axis cs:5.8,0.614)
--(axis cs:5.8,2.616)
--(axis cs:5.2,2.616)
--(axis cs:5.2,0.614)
--cycle;
\addplot [black, forget plot]
table {%
5.5 0.614
5.5 0.607
};
\addplot [black, forget plot]
table {%
5.5 2.616
5.5 4.119
};
\addplot [black, forget plot]
table {%
5.35 0.607
5.65 0.607
};
\addplot [black, forget plot]
table {%
5.35 4.119
5.65 4.119
};
\path [draw=blue, fill=white, postaction={pattern=north west lines, pattern color=blue}]
(axis cs:5.9,0.616)
--(axis cs:6.5,0.616)
--(axis cs:6.5,3.112)
--(axis cs:5.9,3.112)
--(axis cs:5.9,0.616)
--cycle;
\addplot [black, forget plot]
table {%
6.2 0.616
6.2 0.607
};
\addplot [black, forget plot]
table {%
6.2 3.112
6.2 5.107
};
\addplot [black, forget plot]
table {%
6.05 0.607
6.35 0.607
};
\addplot [black, forget plot]
table {%
6.05 5.107
6.35 5.107
};
\path [draw=green, fill=white, postaction={pattern=crosshatch, pattern color=green}]
(axis cs:6.6,0.614)
--(axis cs:7.2,0.614)
--(axis cs:7.2,2.616)
--(axis cs:6.6,2.616)
--(axis cs:6.6,0.614)
--cycle;
\addplot [black, forget plot]
table {%
6.9 0.614
6.9 0.607
};
\addplot [black, forget plot]
table {%
6.9 2.616
6.9 4.117
};
\addplot [black, forget plot]
table {%
6.75 0.607
7.05 0.607
};
\addplot [black, forget plot]
table {%
6.75 4.117
7.05 4.117
};
\path [draw=red, fill=white, postaction={pattern=north east lines, pattern color=red}]
(axis cs:7.3,0.612)
--(axis cs:7.9,0.612)
--(axis cs:7.9,2.614)
--(axis cs:7.3,2.614)
--(axis cs:7.3,0.612)
--cycle;
\addplot [black, forget plot]
table {%
7.6 0.612
7.6 0.607
};
\addplot [black, forget plot]
table {%
7.6 2.614
7.6 3.63
};
\addplot [black, forget plot]
table {%
7.45 0.607
7.75 0.607
};
\addplot [black, forget plot]
table {%
7.45 3.63
7.75 3.63
};
\path [draw=gray, fill=white, postaction={pattern=north west lines, pattern color=gray}]
(axis cs:8,5.607)
--(axis cs:8.6,5.607)
--(axis cs:8.6,11.107)
--(axis cs:8,11.107)
--(axis cs:8,5.607)
--cycle;
\addplot [black, forget plot]
table {%
8.3 5.607
8.3 0.607
};
\addplot [black, forget plot]
table {%
8.3 11.107
8.3 19.11
};
\addplot [black, forget plot]
table {%
8.15 0.607
8.45 0.607
};
\addplot [black, forget plot]
table {%
8.15 19.11
8.45 19.11
};
\addplot [black, opacity=1, mark=asterisk, mark size=4, mark options={solid}, only marks, forget plot]
table {%
8.3 85.607
};
\addplot [semithick, black, mark=x, mark size=3, mark options={solid}, forget plot]
table {%
1 0.609
};
\addplot [semithick, black, mark=x, mark size=3, mark options={solid}, forget plot]
table {%
1.7 0.612
};
\addplot [semithick, black, mark=x, mark size=3, mark options={solid}, forget plot]
table {%
2.4 0.611
};
\addplot [semithick, black, mark=x, mark size=3, mark options={solid}, forget plot]
table {%
3.1 0.609
};
\addplot [semithick, black, mark=x, mark size=3, mark options={solid}, forget plot]
table {%
3.8 4.621
};
\addplot [semithick, black, mark=x, mark size=3, mark options={solid}, forget plot]
table {%
5.5 1.612
};
\addplot [semithick, black, mark=x, mark size=3, mark options={solid}, forget plot]
table {%
6.2 2.107
};
\addplot [semithick, black, mark=x, mark size=3, mark options={solid}, forget plot]
table {%
6.9 1.612
};
\addplot [semithick, black, mark=x, mark size=3, mark options={solid}, forget plot]
table {%
7.6 1.61
};
\addplot [semithick, black, mark=x, mark size=3, mark options={solid}, forget plot]
table {%
8.3 7.609
};
\addplot [black, forget plot]
table {%
0.7 0.609
1.3 0.609
};
\addplot [black, forget plot]
table {%
1.4 0.612
2 0.612
};
\addplot [black, forget plot]
table {%
2.1 0.611
2.7 0.611
};
\addplot [black, forget plot]
table {%
2.8 0.609
3.4 0.609
};
\addplot [black, forget plot]
table {%
3.5 4.621
4.1 4.621
};
\addplot [black, forget plot]
table {%
5.2 1.612
5.8 1.612
};
\addplot [black, forget plot]
table {%
5.9 2.107
6.5 2.107
};
\addplot [black, forget plot]
table {%
6.6 1.612
7.2 1.612
};
\addplot [black, forget plot]
table {%
7.3 1.61
7.9 1.61
};
\addplot [black, forget plot]
table {%
8 7.609
8.6 7.609
};

\draw[red, thick, dashed] 
% shared link delay: TC 6 (1TC5 pkt, 14 TC 6, 70B header): 0.0208ms, TC 5 (14 TC5, 14 TC6, 70B header): 0.0364ms
% Link Talker-TASSW (shared): TC 6: 0.0208ms, TC 5: 0.0364ms
% Link TASSW-SW (trans):      TC 6: 1.36mus,  TC 5: 1.76mus
% Link SW - UE (trans):       TC 6: 1.36mus,  TC 5: 1.76mus
% reported BD 5G: d=0.075ms,  TC 6: 0.575ms,  TC 5: 5.575ms
% Link gnb-NWTT (shared):     TC 6: 0.0208ms, TC 5: 0.0364ms
% Link NWTT-SW (shared):      TC 6: 0.0208ms, TC 5: 0.0364ms
% Link SW-Listener (shared):  TC 6: 0.0208ms, TC 5: 0.0364ms

    (axis cs:0,0.66092) --
    (axis cs:4.0,0.66092)
    node[pos=1, anchor=south east, xshift=-2pt, font=\footnotesize]{};
\draw[red, thick, dashed] 
    (axis cs:5,5.73) --
    (axis cs:9,5.73) 
    node[pos=1, anchor=south east, xshift=-2pt, font=\footnotesize]{};
\end{axis}

\end{tikzpicture}
        \caption{\label{fig:periodic_delay}}
    \end{subfigure}
    \hfill
    % --- Third plot ---
    \begin{subfigure}{0.5\textwidth}
        \centering
        %\resizebox{\textwidth}{!}{\input{figures/periodic_per_rb.tex}}
        % This file was created with tikzplotlib v0.10.1.
\begin{tikzpicture}
\begin{axis}[
height=5cm,
width=5.8cm,
tick align=inside,
tick pos=left,
x grid style={gray},
xmin=-0.541, xmax=4.541,
xtick style={color=black},
trim axis left,
trim axis right,
xtick={0,1,2,3,4},
xticklabels={
        {GF\\MCS\\adaptive}, 
        {GF\\MCS\\5},
        {GF\\MCS\\11},
        {GF\\MCS\\20},
        {Dyn.},
        {Dyn.\\DC-GBR}
    },
xticklabel style={font=\footnotesize, align=center},
y grid style={gray},
ylabel={Average PER [\%]},
ymajorgrids,
ymin=0, ymax=100,
ytick style={color=black},
%width=0.9\linewidth,
ytick style={color=black,font=\footnotesize,},
xticklabel style = {font=\footnotesize},
yticklabel style  = {font=\footnotesize},
xlabel style = {font=\footnotesize},
ylabel style  = {font=\footnotesize},
legend style={
    at={(0.02,0.98)},    % inside top-left corner
    anchor=north west,    % top-left of legend aligned to this point
    legend columns=1,     % vertical stack
    draw=black,           % draw a box around legend
    fill=white,           % optional: background color
    font=\footnotesize,
    /tikz/every node/.append style={inner ysep=1pt, align=left} % left-align entries
}
]
\draw[draw=blue,fill=white,postaction={pattern=north east lines, pattern color=blue}] (axis cs:-0.31,0) rectangle (axis cs:-0.03,4.34526013422433);
\addlegendimage{ybar,ybar legend,draw=blue,fill=white, pattern color=blue,pattern = north west lines,fill opacity=0.9}
\addlegendentry{PER}

\draw[draw=blue,fill=white,postaction={pattern=north east lines, pattern color=blue}] (axis cs:0.69,0) rectangle (axis cs:0.97,2.33282475703111);
\draw[draw=blue,fill=white,postaction={pattern=north east lines, pattern color=blue}] (axis cs:1.69,0) rectangle (axis cs:1.97,5.80229062207849);
\draw[draw=blue,fill=white,postaction={pattern=north east lines, pattern color=blue}] (axis cs:2.69,0) rectangle (axis cs:2.97,26.9199807224291);
\draw[draw=blue,fill=white,postaction={pattern=north east lines, pattern color=blue}] (axis cs:3.69,0) rectangle (axis cs:3.97,87.6477473296766);
\path [draw=black, semithick]
(axis cs:-0.17,4.16266571520601)
--(axis cs:-0.17,4.52785455324266);

\path [draw=black, semithick]
(axis cs:0.83,2.18690411566192)
--(axis cs:0.83,2.4787453984003);

\path [draw=black, semithick]
(axis cs:1.83,5.17581848701898)
--(axis cs:1.83,6.428762757138);

\path [draw=black, semithick]
(axis cs:2.83,24.3994752524659)
--(axis cs:2.83,29.4404861923922);

\path [draw=black, semithick]
(axis cs:3.83,87.2962602542553)
--(axis cs:3.83,87.999234405098);

\addplot [semithick, black, mark=-, mark size=3, mark options={solid}, only marks]
table {%
-0.17 4.16266571520601
0.83 2.18690411566192
1.83 5.17581848701898
2.83 24.3994752524659
3.83 87.2962602542553
};
\addplot [semithick, black, mark=-, mark size=3, mark options={solid}, only marks]
table {%
-0.17 4.52785455324266
0.83 2.4787453984003
1.83 6.428762757138
2.83 29.4404861923922
3.83 87.999234405098
};
\end{axis}

\begin{axis}[
height=5cm,
width=5.8cm,
axis y line=right,
tick align=inside,
x grid style={gray},
xticklabels={
    },
xmin=-0.541, xmax=4.541,
xtick={0,1,2,3,4},
xtick pos=left,
xtick style={color=black},
y grid style={gray},
ylabel={RB Utilization [\%]},
ymin=0, ymax=100,
ytick pos=right,
ytick style={color=black},
yticklabel style={anchor=west,font=\footnotesize},
ytick style={color=black,font=\footnotesize,},
xticklabel style = {font=\footnotesize},
yticklabel style  = {font=\footnotesize},
xlabel style = {font=\footnotesize},
ylabel style  = {font=\footnotesize},
%width=0.9\linewidth,
legend style={
    at={(0.02,0.85)},    % inside top-left corner 
    anchor=north west,    % top-left of legend aligned to this point
    legend columns=1,     % vertical stack
    draw=black,           % draw a box around legend
    fill=white,           % optional: background color
    font=\footnotesize,
    /tikz/every node/.append style={inner ysep=1pt, align=left} % left-align entries
}
]
\draw[draw=orange,fill=white,postaction={pattern=north west lines, pattern color=orange}] (axis cs:0.03,0) rectangle (axis cs:0.31,24.4830292634752);
\addlegendimage{ybar,ybar legend,draw=orange,fill=white,pattern = north east lines, pattern color=orange,fill opacity=0.9}
\addlegendentry{Used RBs}
\draw[draw=orange,fill=white,postaction={pattern=north west lines, pattern color=orange}] (axis cs:1.03,0) rectangle (axis cs:1.31,58.3954129632517);
\draw[draw=orange,fill=white,postaction={pattern=north west lines, pattern color=orange}] (axis cs:2.03,0) rectangle (axis cs:2.31,35.1856671344664);
\draw[draw=orange,fill=white,postaction={pattern=north west lines, pattern color=orange}] (axis cs:3.03,0) rectangle (axis cs:3.31,21.8931077498831);
\draw[draw=orange,fill=white,postaction={pattern=north west lines, pattern color=orange}] (axis cs:4.03,0) rectangle (axis cs:4.31,37.1588881958522);
\path [draw=black, semithick]
(axis cs:0.17,23.9772034244372)
--(axis cs:0.17,24.9888551025133);

\path [draw=black, semithick]
(axis cs:1.17,58.3745769164244)
--(axis cs:1.17,58.4162490100791);

\path [draw=black, semithick]
(axis cs:2.17,35.1730858937326)
--(axis cs:2.17,35.1982483752003);

\path [draw=black, semithick]
(axis cs:3.17,21.8850691913055)
--(axis cs:3.17,21.9011463084606);

\path [draw=black, semithick]
(axis cs:4.17,36.7587497456166)
--(axis cs:4.17,37.5590266460877);

\addplot [semithick, black, mark=-, mark size=3, mark options={solid}, only marks]
table {%
0.17 23.9772034244372
1.17 58.3745769164244
2.17 35.1730858937326
3.17 21.8850691913055
4.17 36.7587497456166
};
\addplot [semithick, black, mark=-, mark size=3, mark options={solid}, only marks]
table {%
0.17 24.9888551025133
1.17 58.4162490100791
2.17 35.1982483752003
3.17 21.9011463084606
4.17 37.5590266460877
};
\end{axis}
\end{tikzpicture}
        \caption{\label{fig:periodic_per}}
    \end{subfigure}
    \caption{Simulation results of periodic simulation scenario. a) E2E delay, b) \ac{PER} and \ac{RB} utilization.}
    \label{fig:three_plots}
\end{figure}
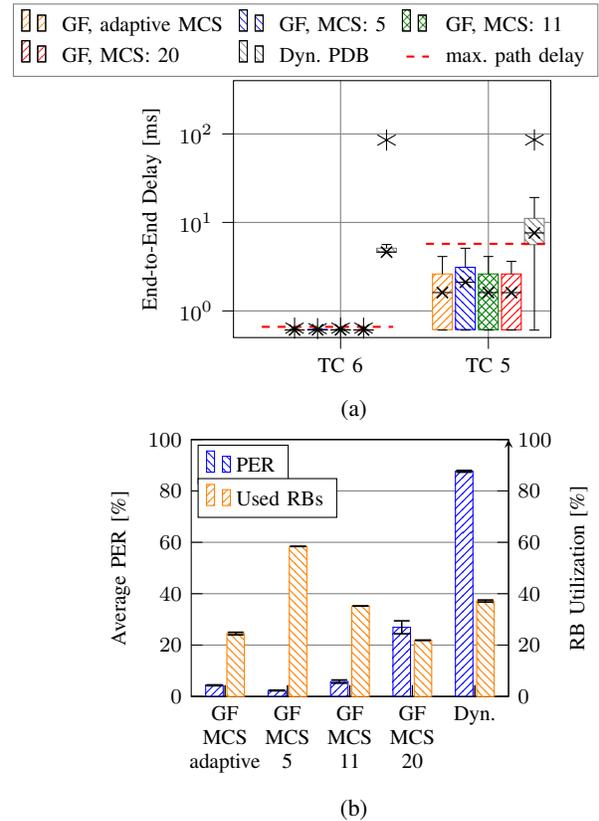

\begin{figure*}[!h]
    \centering
     \begin{tikzpicture}
\definecolor{green}{RGB}{0,128,0} 
\begin{axis}[
    hide axis,           % no axis
    xmin=0, xmax=1,
    ymin=0, ymax=1,
    width=100pt,         % tiny positive width
    height=100pt,        % tiny positive height
    legend style={
        at={(0.3,1)}, anchor=south,
        legend columns=3,    % multi-column legend
        draw=gray,
        column sep=0.9em,
        font=\footnotesize
    },
    legend cell align=left
]

% Define legend symbols manually
\addlegendimage{draw=black, ybar, ybar legend, fill=white, postaction={pattern=north east lines, pattern color=orange}}
\addlegendentry{GF, adaptive MCS + Dyn. Priority (MDBV)}

\addlegendimage{draw=black, ybar, ybar legend, fill=white, postaction={pattern=north west lines, pattern color=blue}}
\addlegendentry{GF, adaptive MCS + Dyn. MAX C/I (MDBV)}

\addlegendimage{draw=black, ybar, ybar legend, fill=white, postaction={pattern=crosshatch, pattern color=green}}
\addlegendentry{GF, adaptive MCS + Dyn. PF (MDBV)}

\addlegendimage{draw=black, ybar, ybar legend, fill=white, postaction={pattern=north east lines, pattern color=red}}
\addlegendentry{GF, adaptive MCS + Dyn. Priority}

\addlegendimage{draw=black, ybar, ybar legend, fill=white, postaction={pattern=north west lines, pattern color=gray}}
\addlegendentry{GF, adaptive MCS + Dyn. MAX C/I}

\addlegendimage{draw=black, ybar, ybar legend, fill=white, postaction={pattern=crosshatch, pattern color=violet}}
\addlegendentry{GF, adaptive MCS + Dyn. PF}

\end{axis}
\end{tikzpicture}
    % --- First plot ---
     \begin{subfigure}[t]{0.32\textwidth}
    % \vspace{0pt}
        \centering
      % \resizebox{\textwidth}{!}{\input{figures/mixed_delay}}
      % This file was created with tikzplotlib v0.10.1.
\begin{tikzpicture}
\definecolor{green}{RGB}{0,128,0}

\begin{axis}[
clip=false,
height=5cm,
width=5.8cm,
enlarge y limits=false,
legend cell align={left},
minor xtick={},
minor ytick={},
log basis y={10},
tick align=outside,
tick pos=left,
x grid style={gray},
xmajorgrids,
xtick={1.7,5.3,8.88},
xmin=-1.05, xmax=10.1,
xticklabels={
{TC 6\\static T\\fixed BS}, {TC 5\\static T\\fixed BS},{TC 4\\sporadic T\\fixed BS}
    },
xticklabel style={font=\small, align=center},									  
y grid style={gray},
ylabel={End-to-End Delay [ms]},
ylabel style={font = \small,},
ymajorgrids,
ymin=0.1, ymax=500,
ymode=log,
ytick={0.1, 1,10,100,1000,10000},
xticklabel style = {font=\small},
yticklabel style  = {font=\small},
xtick style={color=black, font=\small},
ytick style={color=black, font=\small},
yticklabels={
  \(\displaystyle {10^{-1}}\),
  \(\displaystyle {10^{0}}\),
  \(\displaystyle {10^{1}}\),
  \(\displaystyle {10^{2}}\),
  \(\displaystyle {10^{3}}\),
  \(\displaystyle {10^{4}}\)
}  
]
\path [draw=orange, fill=white, postaction={pattern=north east lines, pattern color=orange}]
(axis cs:0.7,0.994)
--(axis cs:1.3,0.994)
--(axis cs:1.3,0.999)
--(axis cs:0.7,0.999)
--(axis cs:0.7,0.994)
--cycle;
\addplot [black, forget plot]
table {%
1 0.994
1 0.992
};
\addplot [black, forget plot]
table {%
1 0.999
1 1.002
};
\addplot [black, forget plot]
table {%
0.85 0.992
1.15 0.992
};
\addplot [black, forget plot]
table {%
0.85 1.002
1.15 1.002
};
\path [draw=blue, fill=white, postaction={pattern=north west lines, pattern color=blue}]
(axis cs:1.4,0.994)
--(axis cs:2,0.994)
--(axis cs:2,0.999)
--(axis cs:1.4,0.999)
--(axis cs:1.4,0.994)
--cycle;
\addplot [black, forget plot]
table {%
1.7 0.994
1.7 0.992
};
\addplot [black, forget plot]
table {%
1.7 0.999
1.7 1.002
};
\addplot [black, forget plot]
table {%
1.55 0.992
1.85 0.992
};
\addplot [black, forget plot]
table {%
1.55 1.002
1.85 1.002
};
\path [draw=green, fill=white, postaction={pattern=crosshatch, pattern color=green}]
(axis cs:2.1,0.994)
--(axis cs:2.7,0.994)
--(axis cs:2.7,0.999)
--(axis cs:2.1,0.999)
--(axis cs:2.1,0.994)
--cycle;
\addplot [black, forget plot]
table {%
2.4 0.994
2.4 0.992
};
\addplot [black, forget plot]
table {%
2.4 0.999
2.4 1.002
};
\addplot [black, forget plot]
table {%
2.25 0.992
2.55 0.992
};
\addplot [black, forget plot]
table {%
2.25 1.002
2.55 1.002
};
\path [draw=orange, fill=white, postaction={pattern=north east lines, pattern color=orange}]
(axis cs:4.3,5.895)
--(axis cs:4.9,5.895)
--(axis cs:4.9,5.904)
--(axis cs:4.3,5.904)
--(axis cs:4.3,5.895)
--cycle;
\addplot [black, forget plot]
table {%
4.6 5.895
4.6 5.893
};
\addplot [black, forget plot]
table {%
4.6 5.904
4.6 5.908
};
\addplot [black, forget plot]
table {%
4.45 5.893
4.75 5.893
};
\addplot [black, forget plot]
table {%
4.45 5.908
4.75 5.908
};
\path [draw=blue, fill=white, postaction={pattern=north west lines, pattern color=blue}]
(axis cs:5,5.895)
--(axis cs:5.6,5.895)
--(axis cs:5.6,5.904)
--(axis cs:5,5.904)
--(axis cs:5,5.895)
--cycle;
\addplot [black, forget plot]
table {%
5.3 5.895
5.3 5.893
};
\addplot [black, forget plot]
table {%
5.3 5.904
5.3 5.908
};
\addplot [black, forget plot]
table {%
5.15 5.893
5.45 5.893
};
\addplot [black, forget plot]
table {%
5.15 5.908
5.45 5.908
};
\path [draw=green, fill=white, postaction={pattern=crosshatch, pattern color=green}]
(axis cs:5.7,5.895)
--(axis cs:6.3,5.895)
--(axis cs:6.3,5.904)
--(axis cs:5.7,5.904)
--(axis cs:5.7,5.895)
--cycle;
\addplot [black, forget plot]
table {%
6 5.895
6 5.893
};
\addplot [black, forget plot]
table {%
6 5.904
6 5.908
};
\addplot [black, forget plot]
table {%
5.85 5.893
6.15 5.893
};
\addplot [black, forget plot]
table {%
5.85 5.908
6.15 5.908
};
\path [draw=orange, fill=white, postaction={pattern=north east lines, pattern color=orange}]
(axis cs:7.9,6.227)
--(axis cs:8.5,6.227)
--(axis cs:8.5,9.069)
--(axis cs:7.9,9.069)
--(axis cs:7.9,6.227)
--cycle;
\addplot [black, forget plot]
table {%
8.2 6.227
8.2 1.969
};
\addplot [black, forget plot]
table {%
8.2 9.069
8.2 13.332
};
\addplot [black, forget plot]
table {%
8.05 1.969
8.35 1.969
};
\addplot [black, forget plot]
table {%
8.05 13.332
8.35 13.332
};
\addplot [black, opacity=1, mark=asterisk, mark size=2.5, mark options={solid}, only marks, forget plot]
table {%
8.2 61.555
};
\path [draw=blue, fill=white, postaction={pattern=north west lines, pattern color=blue}]
(axis cs:8.6,6.261)
--(axis cs:9.2,6.261)
--(axis cs:9.2,9.125)
--(axis cs:8.6,9.125)
--(axis cs:8.6,6.261)
--cycle;
\addplot [black, forget plot]
table {%
8.9 6.261
8.9 1.97
};
\addplot [black, forget plot]
table {%
8.9 9.125
8.9 13.421
};
\addplot [black, forget plot]
table {%
8.75 1.97
9.05 1.97
};
\addplot [black, forget plot]
table {%
8.75 13.421
9.05 13.421
};
\addplot [black, opacity=1, mark=asterisk, mark size=2.5, mark options={solid}, only marks, forget plot]
table {%
8.9 222.928
};
\path [draw=green, fill=white, postaction={pattern=crosshatch, pattern color=green}]
(axis cs:9.3,6.236)
--(axis cs:9.9,6.236)
--(axis cs:9.9,9.092)
--(axis cs:9.3,9.092)
--(axis cs:9.3,6.236)
--cycle;
\addplot [black, forget plot]
table {%
9.6 6.236
9.6 1.955
};
\addplot [black, forget plot]
table {%
9.6 9.092
9.6 13.376
};
\addplot [black, forget plot]
table {%
9.45 1.955
9.75 1.955
};
\addplot [black, forget plot]
table {%
9.45 13.376
9.75 13.376
};
\addplot [black, opacity=1, mark=asterisk, mark size=2.5, mark options={solid}, only marks, forget plot]
table {%
9.6 174.981
};
\addplot [line width=0.48pt, black, mark=x, mark size=3, mark options={solid}, forget plot]
table {%
1 0.996
};
\addplot [line width=0.48pt, black, mark=x, mark size=3, mark options={solid}, forget plot]
table {%
1.7 0.996
};
\addplot [line width=0.48pt, black, mark=x, mark size=3, mark options={solid}, forget plot]
table {%
2.4 0.996
};
\addplot [line width=0.48pt, black, mark=x, mark size=3, mark options={solid}, forget plot]
table {%
4.6 5.9
};
\addplot [line width=0.48pt, black, mark=x, mark size=3, mark options={solid}, forget plot]
table {%
5.3 5.9
};
\addplot [line width=0.48pt, black, mark=x, mark size=3, mark options={solid}, forget plot]
table {%
6 5.9
};
\addplot [line width=0.48pt, black, mark=x, mark size=3, mark options={solid}, forget plot]
table {%
8.2 7.65
};
\addplot [line width=0.48pt, black, mark=x, mark size=3, mark options={solid}, forget plot]
table {%
8.9 7.689
};
\addplot [line width=0.48pt, black, mark=x, mark size=3, mark options={solid}, forget plot]
table {%
9.6 7.669
};
\addplot [black, forget plot]
table {%
0.7 0.996
1.3 0.996
};
\addplot [black, forget plot]
table {%
1.4 0.996
2 0.996
};
\addplot [black, forget plot]
table {%
2.1 0.996
2.7 0.996
};
\addplot [black, forget plot]
table {%
4.3 5.9
4.9 5.9
};
\addplot [black, forget plot]
table {%
5 5.9
5.6 5.9
};
\addplot [black, forget plot]
table {%
5.7 5.9
6.3 5.9
};
\addplot [black, forget plot]
table {%
7.9 7.65
8.5 7.65
};
\addplot [black, forget plot]
table {%
8.6 7.689
9.2 7.689
};
\addplot [black, forget plot]
table {%
9.3 7.669
9.9 7.669
};
\node[anchor=south west, font=\small] at (rel axis cs:0.42,-0.5) {(a)};
\end{axis}

\end{tikzpicture}
      \captionsetup{labelformat=empty}
      \caption{}
       \label{fig:delay_mixed}
    \end{subfigure}
    \hfill
        \begin{subfigure}[t]{0.32\textwidth}
   % \vspace{0pt}
        \centering
      %  \resizebox{\textwidth}{!}{\input{figures/TC4_Delay.tex}}
      % This file was created with tikzplotlib v0.10.1.
\begin{tikzpicture}

\definecolor{green}{RGB}{0,128,0}

\begin{axis}[
height=5cm,
width=5.7cm,
enlarge y limits=false,
log basis y={10},
tick align=outside,
tick pos=left,
x grid style={gray},
xmajorgrids,
xtick={0,1,2},				 
xmin=-0.496, xmax=2.496,
xticklabels={
{Mean}, {99th\\Percentile}, {99.9th\\Percentile}
	},
xticklabel style={font=\small, align=center},
y grid style={gray},
ylabel={End-to-End Delay [ms]},
ylabel style={font = \small,},
ymajorgrids,
ymin=0.1, ymax=800,
ymode=log,
clip=false,		
ytick={0.1,1,10,100},
xticklabel style = {font=\small},
yticklabel style  = {font=\small},
xtick style={color=black, font=\small},
ytick style={color=black, font=\small},
yticklabels={
  \(\displaystyle {10^{-1}}\),
  \(\displaystyle {10^{0}}\),
  \(\displaystyle {10^{1}}\),
  \(\displaystyle {10^{2}}\),
  \(\displaystyle {10^{3}}\),
  \(\displaystyle {10^{4}}\)
}  
]
\draw[draw=orange,fill=white,line width=0.32pt,postaction={pattern=north east lines, pattern color=orange}] (axis cs:-0.36,0.1) rectangle (axis cs:-0.26,7.95136176840367);
%\addlegendimage{ybar,ybar legend,draw=orange,fill=white,line width=0.32pt,postaction={pattern=north east lines, pattern color=orange}}
%\addlegendentry{Priority_TC4}

\draw[draw=orange,fill=white,line width=0.32pt,postaction={pattern=north east lines, pattern color=orange}] (axis cs:0.64,0.1) rectangle (axis cs:0.74,14.965);
\draw[draw=orange,fill=white,line width=0.32pt,postaction={pattern=north east lines, pattern color=orange}] (axis cs:1.64,0.1) rectangle (axis cs:1.74,21.751736000001);
\draw[draw=blue,fill=white,line width=0.32pt,postaction={pattern=north west lines, pattern color=blue}] (axis cs:-0.24,0.1) rectangle (axis cs:-0.14,8.66915136325805);
%\addlegendimage{ybar,ybar legend,draw=blue,fill=white,line width=0.32pt,postaction={pattern=north west lines, pattern color=blue}}
%\addlegendentry{MAXCI_TC4}

\draw[draw=blue,fill=white,line width=0.32pt,postaction={pattern=north west lines, pattern color=blue}] (axis cs:0.76,0.1) rectangle (axis cs:0.86,33.947);
\draw[draw=blue,fill=white,line width=0.32pt,postaction={pattern=north west lines, pattern color=blue}] (axis cs:1.76,0.1) rectangle (axis cs:1.86,84.8044000000011);
\draw[draw=green,fill=white,line width=0.32pt,postaction={pattern=crosshatch, pattern color=green}] (axis cs:-0.12,0.1) rectangle (axis cs:-0.02,8.33283542950061);
%\addlegendimage{ybar,ybar legend,draw=green,fill=white,line width=0.32pt,postaction={pattern=crosshatch, pattern color=green}}
%\addlegendentry{PF_TC4}

\draw[draw=green,fill=white,line width=0.32pt,postaction={pattern=crosshatch, pattern color=green}] (axis cs:0.88,0.1) rectangle (axis cs:0.98,24.72702);
\draw[draw=green,fill=white,line width=0.32pt,postaction={pattern=crosshatch, pattern color=green}] (axis cs:1.88,0.1) rectangle (axis cs:1.98,46.9897560000008);
\draw[draw=red,fill=white,line width=0.32pt,postaction={pattern=north east lines, pattern color=red}] (axis cs:-6.93889390390723e-18,0.1) rectangle (axis cs:0.1,8.89398512699069);
%\addlegendimage{ybar,ybar legend,draw=red,fill=white,line width=0.32pt,postaction={pattern=north east lines, pattern color=red}}
%\addlegendentry{Priority}

\draw[draw=red,fill=white,line width=0.32pt,postaction={pattern=north east lines, pattern color=red}] (axis cs:1,0.1) rectangle (axis cs:1.1,16.80849);
\draw[draw=red,fill=white,line width=0.32pt,postaction={pattern=north east lines, pattern color=red}] (axis cs:2,0.1) rectangle (axis cs:2.1,23.4731280000005);
\draw[draw=gray,fill=white,line width=0.32pt,postaction={pattern=north west lines, pattern color=gray}] (axis cs:0.12,0.1) rectangle (axis cs:0.22,14.0136412375132);
%\addlegendimage{ybar,ybar legend,draw=gray,fill=white,line width=0.32pt,postaction={pattern=north west lines, pattern color=gray}}
%\addlegendentry{MAXCI}

\draw[draw=gray,fill=white,line width=0.32pt,postaction={pattern=north west lines, pattern color=gray}] (axis cs:1.12,0.1) rectangle (axis cs:1.22,126.007960000001);
\draw[draw=gray,fill=white,line width=0.32pt,postaction={pattern=north west lines, pattern color=gray}] (axis cs:2.12,0.1) rectangle (axis cs:2.22,606.017172000037);
\draw[draw=violet,fill=white,line width=0.32pt,postaction={pattern=crosshatch, pattern color=violet}] (axis cs:0.24,0.1) rectangle (axis cs:0.34,10.4579607887296);
%\addlegendimage{ybar,ybar legend,draw=violet,fill=white,line width=0.32pt,postaction={pattern=crosshatch, pattern color=violet}}
%\addlegendentry{PF}

\draw[draw=violet,fill=white,line width=0.32pt,postaction={pattern=crosshatch, pattern color=violet}] (axis cs:1.24,0.1) rectangle (axis cs:1.34,39.154);

\draw[draw=violet,fill=white,line width=0.32pt,postaction={pattern=crosshatch, pattern color=violet}] (axis cs:2.24,0.1) rectangle (axis cs:2.34,94.4805510000015);
table {%
-0.45 50
2.475 50
};

\draw[red, thick, dashed] 
    (axis cs:-0.45,50) --
    (axis cs:2.475,50) 
    node[pos=1, anchor=south east, xshift=-2pt, font=\large]{};

\draw (axis cs:-0.47,30) node[
  scale=0.7,
  anchor=south east,
  text=red,
  rotate=0.0
]{PDB};

\node[anchor=south west, font=\small] at (rel axis cs:0.42,-0.5) {(b)};
\end{axis}

\end{tikzpicture}
      \captionsetup{labelformat=empty}
      \caption{}
      \label{fig:delayTC4}
    \end{subfigure}
    % --- Third plot ---
    \begin{subfigure}[t]{0.32\textwidth}
    %\vspace{0pt}
        \centering
      %  \resizebox{\textwidth}{!}{\input{figures/tp_cv}}
      \input{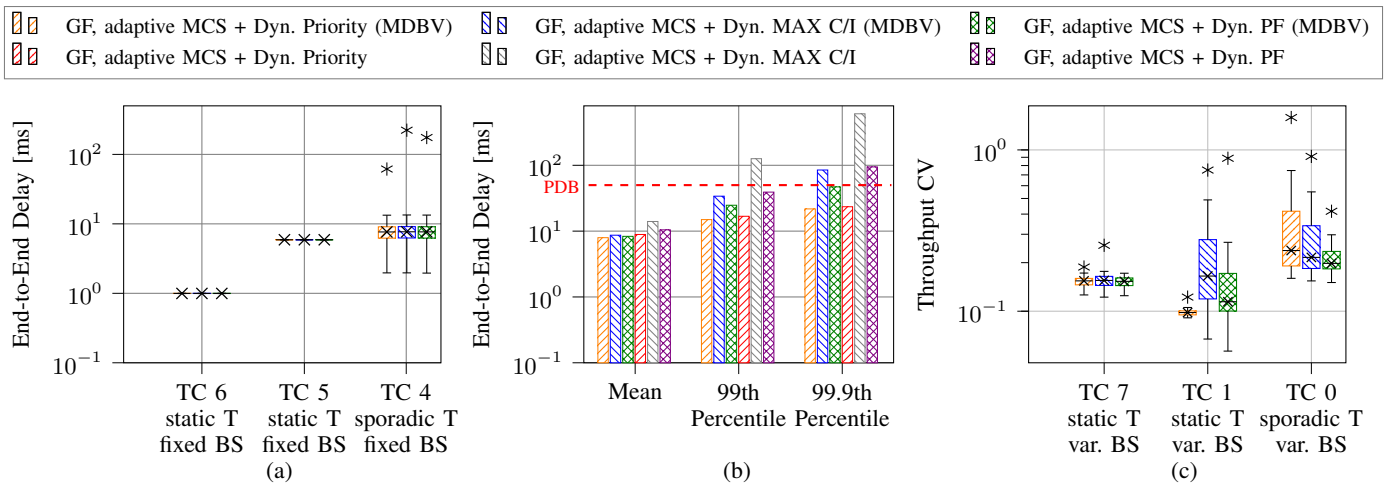}
      \captionsetup{labelformat=empty}
      \caption{}
    \label{fig:tp}
    \end{subfigure}
    \hfill
    % --- Second plot ---
    \vspace{-1cm}
    \caption{Results of heterogeneous simulation scenario. a) E2E delay of time-sensitive \acp{TC}, b) Delay and jitter of non-deterministic time-sensitive flows showing impact of the \ac{MDBV} configuration, and c) Throughput stability of \acp{TC} with a rate requirement.}
    \label{fig:three_plots}
\end{figure*}
%%The periodic scenario evaluates the grant-free static allocation capabilities of our proposed heterogeneous scheduler. We compare our approach to pre-allocate resources for periodic flows, but to adjust the \ac{MCS} and allocation  in case of changing channel conditions to two baselines. The first baseline uses a static schedule with a fixed \ac{MCS} following the principle of \ac{CG} Comparing different static \ac{MCS} configurations enables us to consider the full capabilities of the \ac{CG} approach, ensuring a fair comparison to our proposed scheduler.
% Although the algorithm of is not implemented due to the missing applicability for \ac{TDD} patterns, \cite{Larrañaga} relies on static \ac{MCS} selection such that the scheduling based on a fixed \ac{MCS} is used to compare the proposed scheduler to \cite{Larrañaga} in terms of reliability and resource efficiency. 
%As a second baseline, the dynamic resource scheduler of~\cite{Debnath} is considered, prioritizing flows based on their \ac{PDB}. According to the definition of \ac{DC-GBR} flows in 5G~\cite{23.501}, packets exceeding their \ac{PDB} on the \ac{RLC} layer shall be dropped. To consider both, the scheduler is simulated once with and without dropping based on the delay.

\subsection{Grant-Free Scheduling of Time-Sensitive Flows}
In the periodic scenario, we focus on the scheduling of periodic time-sensitive \acp{TC} to evaluate our grant-free scheduling algorithm with respect to end-to-end delay, resource utilization, and \ac{PER}. The following  schedulers are compared: 
\begin{enumerate}
    \item Our proposed grant-free allocation scheme that dynamically adjusts the \ac{MCS} to the channel.
    \item Grant-free allocation based on 3 different static \ac{MCS} to enable comparison with \acp{CG} as defined by 3GPP~\cite{38.321, 38.331} and used in~\cite{Larrañaga}.
    \item \ac{PDB}-based dynamic resource scheduling as introduced by~\cite{Debnath}. Packets exceeding their \ac{PDB} at the \ac{RLC} layer are dropped, in line with the definition of \ac{DC-GBR} flows in 5G~\cite{23.501}.% The scheduler is simulated once with and without dropping based on the \ac{PDB}. 
\end{enumerate}

Fig.~\ref{fig:periodic_delay} visualizes the end-to-end delay in the periodic simulation scenario, highlighting that only the grant-free schedulers guarantee the configured maximum path delay, defined as the sum of the reported maximum \acp{BD}. While the dynamic scheduler drops \ac{DC-GBR} frames that exceed their deadlines from the \ac{RLC} queue, this does not prevent frames from experiencing high delays at the lower MAC layer, highlighting the need for a grant-free scheduler. 
Comparing the end-to-end delays of both periodic \acp{TC} under grant-free scheduling illustrates the different configured \ac{QoS} levels: \ac{TC} 5 experiences jitter in the range of the minimum and maximum \ac{BD}, whereas \ac{TC} 6 is transmitted with almost no jitter. 

Fig.~\ref{fig:periodic_per} shows the \ac{PER} and resource efficiency, where \ac{PER} is the ratio of packets received within the maximum reported path delay to the total number of sent packets and resource efficiency is the ratio of used \acp{RB} to the total number of schedulable \acp{RB} in the \ac{UL}. 
Our proposed scheme outperforms the dynamic scheduler in terms of reliability, due to the lack of missing real-time scheduling in the latter. In addition, it outperforms the \ac{CG} baseline by significantly improving the resource efficiency while maintaining comparable reliability. %Concluding, the results demonstrate that our grant-free allocation scheme guarantees the configured bridge delay while balancing reliability and resource efficiency. 

\subsection{E2E Scheduling of TAS and Heterogeneous \ac{RAN} Scheduler}
The heterogeneous scenario simulates a typical 5G industrial use case, involving \acp{TC} with diverse traffic characteristics. The 5G link is integrated as a transparent bridge within a chain of \ac{TSN} switches, as shown in Table~\ref{tab:sim}. To ensure the deterministic transmission of flows belonging to the time-critical \acp{TC} 5 and 6 alongside non-deterministic flows, a \ac{TAS} schedule is applied. The first switch SW1 ensures deterministic frame arrival at the 5G bridge, while the schedule at the second switch SW2 is adjusted by shifting dedicated slots according to the maximum 5G \ac{BD} of each \ac{TC}. To reduce the complexity associated with managing multiple \ac{PDU} sessions per \ac{UE}, each \ac{UE} is configured to transmit a single stream of data. To serve non-deterministic flows, our heterogeneous \ac{RAN} scheduler combines the proposed grant-free allocation scheme with the following dynamic schedulers:
\begin{enumerate*}
    \item Priority-based scheduler proposed by~\cite{Debnath}
    \item \ac{PF} 
    \item Max C/I.
\end{enumerate*}

Fig.~\ref{fig:delay_mixed} shows the end-to-end delay of the time-sensitive \acp{TC} in the heterogeneous simulation scenario. 
Similar to the periodic simulation scenario, periodic time-sensitive flows achieve low latency due to grant-free scheduling in the 5G link. However, \ac{TC} 5 experiences less jitter because the \ac{TAS} schedule de-jitters the flows after the 5G link at SW2, showing the end-to-end scheduling capabilities. 

The average delay of the non-deterministic, time-sensitive \ac{TC}~4 remains below 10~\si{ms}, even though it is scheduled dynamically. 
This is achieved through the \ac{MDBV} configuration, which supplies the scheduler with the maximum \ac{BS} of \ac{TC} 4 flows, allowing it to allocate enough resources in the initial grant to transmit the entire packet and thereby reducing overall delay. Fig.~\ref{fig:delayTC4} shows this effect by comparing the scheduler performance with and without the configured \ac{MDBV} parameter. Across all simulated dynamic scheduling disciplines, the probability of exceeding the \ac{PDB} of 50~\si{ms} (equal to its maximum periodicity) decreases noticeably when the \ac{MDBV} parameter is configured. Among the tested approaches, priority-based scheduling achieves the lowest jitter due to the prioritization of the \ac{TC} in the retransmission case.

The coefficient of variation (CV) is used in Fig.~\ref{fig:tp} to quantify the stability of throughput. By calculating the CV on 1~\si{s} average throughput intervals for each \ac{UE} and simulation run, we can compare the stability of different \acp{TC} and schedulers, even when their absolute throughput levels differ. Fig.~\ref{fig:tp} shows that the priority-based scheduling provides the best throughput stability for non-best effort \acp{TC}, but \ac{PF} ensures a fairly low throughput variation for all \acp{TC} such that throughput requirements have to be taken into account to decide which dynamic scheduling strategy is suited best. 
\section{Conclusion} \label{sec:conclusion}
This paper introduces a joint scheduling approach to integrate a 5G \ac{UL} as a transparent bridge with time-aware \ac{TSN}. To enable end-to-end scheduling across heterogeneous transmission domains, the 5G \ac{BD} is analyzed and calculated for each \ac{TC}, classifying traffic into statically and dynamically scheduled flows. The static radio resource allocation ensures a conflict-free schedule and dynamically adjusts the \ac{MCS} according to the current channel conditions, thereby enhancing reliability in mobility scenarios.
The performance analysis of the proposed radio resource scheduler demonstrates its ability to guarantee various \ac{QoS} levels and to jointly schedule time-sensitive flows across the wired and wireless domains. By accounting for \ac{TDD} patterns and control overhead for \ac{UL} scheduling, the scheduler is applicable to real-world testbeds. The integration in a 5G-\ac{TSN} testbed, such as \cite{CnsmDemo, Rodriguez}, is part of future work. Further research could focus on optimizing the scheduling process by reducing the number of rescheduling events that adjust the grant-free allocated resources. Additionally, incorporating target reliability into scheduling decisions could enhance performance, not only to schedule streams on time but also to guarantee the defined \ac{PER}. 

%Additionally, aligning the dynamic scheduling of streams requiring a target throughput with the CBS of the \ac{TSN} domain could improve throughput stability, as proposed in~\cite{cbs}.

\bibliography{literature.bib}
\bibliographystyle{IEEEtran}

\end{document}